\documentclass[a4paper]{jpconf}
\usepackage{graphicx}
\usepackage{amsmath}
\begin{document}
\title{Dissipation in PIC simulations of moderate to low $\beta$ plasma turbulence}

\author{Kirit Makwana$^{1}$, Hui Li$^2$, Fan Guo$^2$ and Xiaocan Li$^3$}

\address{$^1$ Center for mathematical Plasma Astrophysics, KU Leuven, Belgium}
\address{$^2$ Los Alamos National Laboratory, New Mexico, USA}
\address{$^3$ University of Alabama-Huntsville, Alabama, USA}

\ead{kirit.makwana@kuleuven.be}

\begin{abstract}
We simulate decaying turbulence in electron-positron pair plasmas using a fully-kinetic particle-in-cell (PIC) code. We run two simulations with moderate-to-low  plasma $\beta$. The energy decay rate is found to be similar in both the cases. The perpendicular wave-number spectrum of magnetic energy shows a slope of $k_{\perp}^{-1.3}$ in both the cases. The particle energy distribution function shows the formation of a non-thermal feature in the case of lower plasma $\beta$, with a slope close to $E^{-1}$. The role of thin turbulent current sheets in this process is investigated. The heating by $\bf{E}_{\parallel}\cdot\bf{J}_{\parallel}$ term dominates the $\bf{E}_{\perp}\cdot\bf{J}_{\perp}$ term. Regions of strong $\bf{E}_{\parallel}\cdot\bf{J}_{\parallel}$ are spatially well-correlated with regions of intense current sheets, which also appear correlated with regions of strong $\bf{E}_{\parallel}$ in the low $\beta$ simulation, suggesting an important role of magnetic reconnection in the dissipation of low $\beta$ plasma turbulence.
\end{abstract}

\section{Introduction}

Plasma turbulence is ubiquitous throughout space and astrophysical systems. Turbulence in the solar corona and solar wind is an important candidate mechanism for explaining their heating~\cite{LeamonMatthaeus2000}. Turbulence can also be a possible source of heat in intracluster medium to balance radiative cooling~\cite{ZhuravlevaChurazov2014}. Alfven waves in the solar wind~\cite{BelcherDavis1971} nonlinearly interact with each other and produce turbulence. 3D turbulence is characterized by an energy cascade from large to small scales~\cite{Kolmogorov1941}. At small enough scales energy dissipation takes place. In plasmas which are nearly collision-less, kinetic mechanisms have to operate to dissipate turbulent energy. Recent observations of solar wind show the energy cascade even extending from the ion scales to the electron scales~\cite{AlexandrovaSaur2009}. At the same time, current sheets are also observed in the solar wind~\cite{Li2008}. These current sheets can also dissipate energy through the mechanism of magnetic reconnection~\cite{Karimabadi2013}. Therefore it is necessary to understand the correct energy dissipation mechanisms in such turbulence, and for this numerical simulations are needed.

We have simulated decaying plasma turbulence at moderate plasma $\beta$ (ratio of thermal to magnetic pressure) using both MHD and PIC simulations previously~\cite{makwana2015}. In that study we found that the simulations compare remarkably well. The turbulent power spectra were well-resolved and for large enough simulation box sizes, the spectral slope matched very well between MHD and PIC simulations. This showed that PIC simulations can reproduce MHD results at large scales. Both simulations showed formation of thin current sheets, which are also observed in solar wind plasmas~\cite{Li2008}. The lengths of the current sheets scaled linearly with the driving scale of turbulence, in both MHD and PIC. On the other hand, we saw important differences at small scales. We measured the dissipation fraction as a function of the volume fraction of current sheets. We found that PIC simulations showed more diffuse current sheets compared to MHD. Explicitly measuring the thickness of current sheets revealed that they are of the order of skin-depth scale in PIC whereas they are primarily set by the grid spacing in MHD. This showed that the PIC simulations were successfully capturing the whole range of physics from MHD to kinetic scales. 

These simulations were performed at a moderate plasma $\beta=0.33$. We also observed particle heating in these simulations as the turbulence decays. Recent PIC simulations of magnetic reconnection at low plasma $\beta$ have revealed strong signatures of non-thermal particle heating~\cite{LiGuo2015}. The plasma $\beta$ can vary significantly in the solar corona, from $\sim 10^{-3}$ at the base of corona to $\sim 10$ at the top~\cite{Aschwanden}. At the same time, stochastic acceleration of particles in turbulence is considered an important acceleration mechanism in a  variety of astrophysical sources, from solar flares to galaxy clusters~\cite{Petrosian2012}. Therefore it is important to understand particle heating and acceleration by turbulence at varying plasma $\beta$. We present results of a couple of PIC simulations of decaying turbulence with moderate to low plasma $\beta$. Sec. 2 describes the setup of the simulations, their energy dynamics and the turbulent power spectra, Sec. 3 shows the particle energization and an analysis of the energization terms along with the role of accompanying current sheets. We conclude with discussion in Sec. 4.

\section{Setup of simulations and energy spectra}
We simulate decaying turbulence by setting up an ensemble of waves as the initial condition in the simulation box. These waves are Alfvenic in nature with the characteristic that the magnetic and velocity perturbations are perpendicular to the background field and are parallel/anti-parallel, with no density or pressure perturbations. The form of these perturbations is given in Eqns. (1) and (2) of Ref.~\cite{makwana2015}. The thermal velocity is $v_{th,i,e}=0.08c$. The plasma is an electron-positron pair plasma with equal ion-electron masses. Pair plasmas are believed to be emitted in astrophysical compact objects like blackholes~\cite{BegelmanBlandford1984} and pulsars~\cite{GoldreichJulian1969}. There is also a need to explain non-thermal particle acceleration in pair-plasmas in pulsars, active galactic nuclei, and gamma-ray-bursts~\cite{HoshinoLyubarsky2012}. Therefore our simulations are relevant for such problems. The particles are initialized with a Maxwellian distribution around a mean velocity defined by the linear combination of the fluid velocity and current density. Since it is a pair-plasma, the electrons and ions both contribute equally to the current density and flow. The ratio of plasma frequency to cyclotron frequency, $\omega_{p,i}/\omega_{c,i}$, sets the ion $\beta$ which is given by $\beta_{i}=2(\omega_{p,i}/\omega_{c,i})^2(v_{th,i}/c)^2$. For pair plasmas, the constants for ions and electrons are same and therefore the plasma $\beta$ is simply twice this value. In Ref.~\cite{makwana2015}, all the simulations had $\beta=0.33$. In this study we have two cases, case 1 with $\beta=0.092$ and case 2 with $\beta=0.026$. As we vary the plasma $\beta$, we keep the ratio of injected energy (in the waves) to background magnetic field energy at a constant level of around 0.2. The simulations are performed with the state-of-the-art PIC code VPIC~\cite{BowersAlbright2008}.

\begin{figure}[h]
\includegraphics[width=21pc]{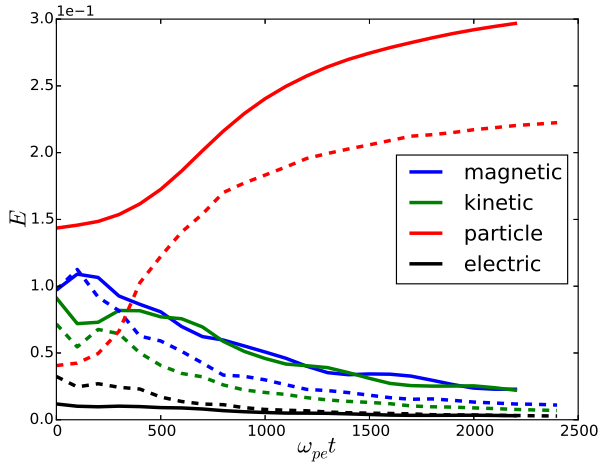}\hspace{1pc}%
\begin{minipage}[b]{13pc}\caption{\label{energy}Time evolution of energy. Solid lines are for higher $\beta=0.092$ case while dashed lines are for the lower $\beta=0.026$ case. The magnetic energy only includes the perpendicular magnetic field, not the background field. The particle energy is the thermal energy, without the kinetic component. The energy is normalized w.r.t. background magnetic field energy. That is why the thermal energy in lower $\beta$ run is lower.}
\end{minipage}
\end{figure}
The dimensions of the simulation box are $(120d_i,120d_i,480d_i)$, where $d_i$ is the ion skin depth, with a resolution of 576 cells in each direction. The simulations are run for at least $2000\omega_{p,i}^{-1}$ time. The setup of initial waves is balanced i.e., there is equal energy in waves moving in opposite directions. We start from an initial state which is already in critical balance with $k_{\perp}\delta b_{\perp}\sim k_{\parallel}B_0$~\cite{Goldreich1995}. As these waves interact non-linearly, they generate turbulence, the energy cascades forward to smaller scales, there it dissipates and the energy in the waves decays, going into particle energy. The total energy is very well conserved in these simulations, with the finite-grid heating leading to an increase of less than 0.2\% increase in the total energy. We look at the decay of energy in Fig.~\ref{energy}. Since the plasma $\beta$ changes by changing the background magnetic field, we normalize all energies by the background magnetic field. Thus, we can see similar level of magnetic and kinetic energy in the waves at initial time for both cases. Naturally the thermal energy is lower for the lower plasma $\beta$ case. For $\omega_{pe}t < 200$ we can see increase of magnetic energy and decrease of kinetic energy. This is consistent with our earlier simulations~\cite{makwana2015}. The kinetic energy is smaller than magnetic energy in lower $\beta$ case because the relativistic Alfven velocity becomes smaller compared to background magnetic field. The decay rate of kinetic and magnetic energy look very similar in both the cases, as also the heating rate of the particles. We see a similar amount of thermal energy being generated in both the cases.



\begin{figure}[h]
\center
\includegraphics[width=1.0\textwidth]{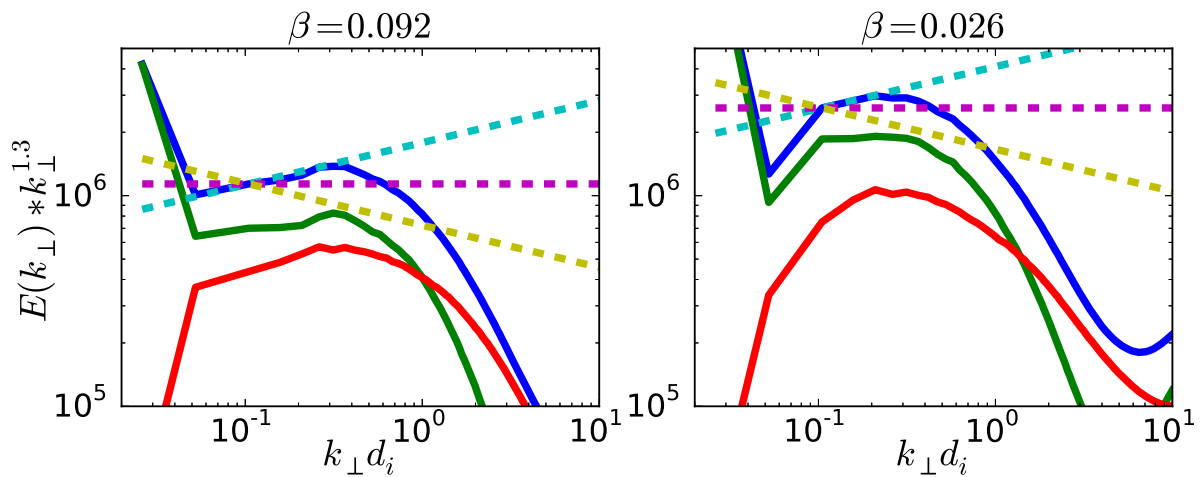}\hspace{2pc}%
\caption{\label{spectrum}The perpendicular wavenumber spectrum for the two cases. Blue curve is total energy, green is magnetic energy and red is kinetic energy spectrum. The spectrum is normalized by $k_{\perp}^{1.3}$, so the dashed, horizontal magenta line is a slope of $k_{\perp}^{-1.3}$, the yellow dashed line is $k_{\perp}^{-1.5}$, and the blue dashed line is $k_{\perp}^{-1.1}$ slope.}
\end{figure}
The power-spectrum of turbulence also decays with energy, however, it shows a steady power-law slope while it decays, which we average from $\omega_{p,e}t=1000$ to $\omega_{p,e}t=2200$ by compensating for the decay with time. We show this time-averaged perpendicular wave-number spectrum in the Fig.~\ref{spectrum}. The cascade in parallel wave-number is weak. The spectrum is multiplied by $k_{\perp}^{1.3}$ in order to compare with previous results in Ref.~\cite{makwana2015}. In the previous simulations at $\beta=0.33$ at this resolution the spectral slope was closer to $k_{\perp}^{-1.5}$ (Fig.3f in Ref.~\cite{makwana2015}). In these simulations, the magnetic energy is showing a slope close to $k_{\perp}^{-1.3}$. For the earlier $\beta=0.33$ runs, the spectral slope tended to $k_{\perp}^{-1.3}$ when the simulation box was made larger, matching the spectral slope in MHD simulations, whereas in these cases we are already getting the MHD slope at a smaller box size. The kinetic energy spectrum does not show a clear inertial range. We see that the spectrum turns over close to skin-depth scale $k_{\perp}d_i\approx1.0$, also where the magnetic energy spectrum dips below the kinetic energy spectrum.


\section{Particle energization and $\mathbf{E}\cdot\mathbf{J}$ analysis}
\begin{figure}[h]
\center
\includegraphics[width=1.0\textwidth]{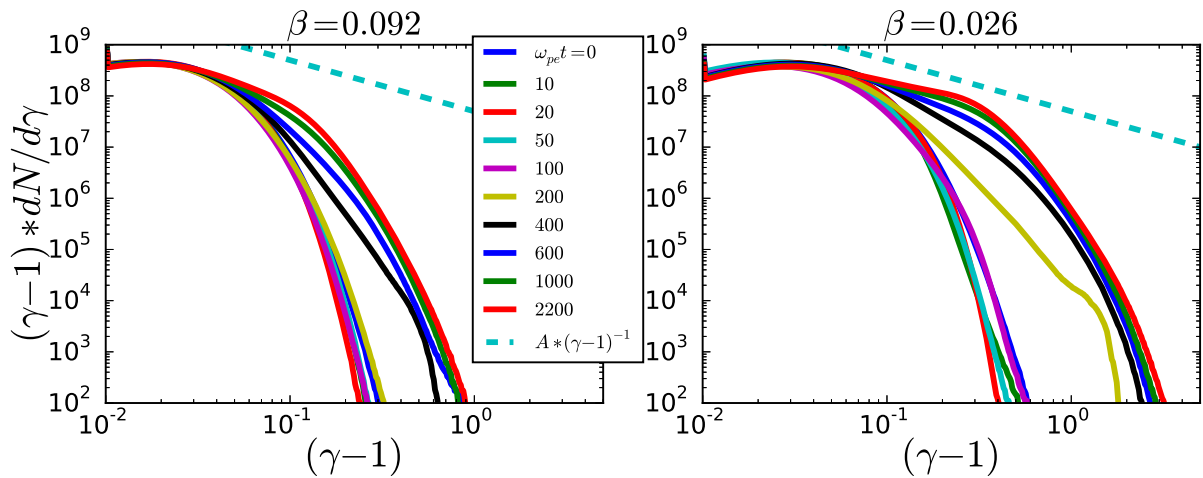}\hspace{2pc}%
\caption{\label{energy_dist}Distribution function of kinetic energy of particles at various times for the 2 cases. The time is normalized in units of $\omega_{pe}^{-1}$. The dashed line is at a slope of $(\gamma -1)^{-1}$.}
\end{figure}

We look at the time evolution of the distribution of particle kinetic energy in these simulations in Fig.~\ref{energy_dist}. The initial distribution is Maxwellian, and the particles are energized to higher energy as the turbulence heats them. By $\omega_{p,e}t=1000$ the distribution has become steady. For the $\beta=0.092$ case we see that the distribution function extends to higher energies as the particles are heated, while keeping a Maxwellian shape. For $\beta=0.026$ case, the distribution develops a flat, non-thermal feature which extends up to $(\gamma-1)\approx0.3$, i.e. energies of up to 150 keV for electrons and positrons. The slope of this feature is close to $(\gamma-1)^{-1}$. This feature begins to appear at around $\omega_{pe}t=200$ and reaches a steady level by around $\omega_{pe}t=1000$.
 
\begin{figure}[h]
\includegraphics[width=24pc]{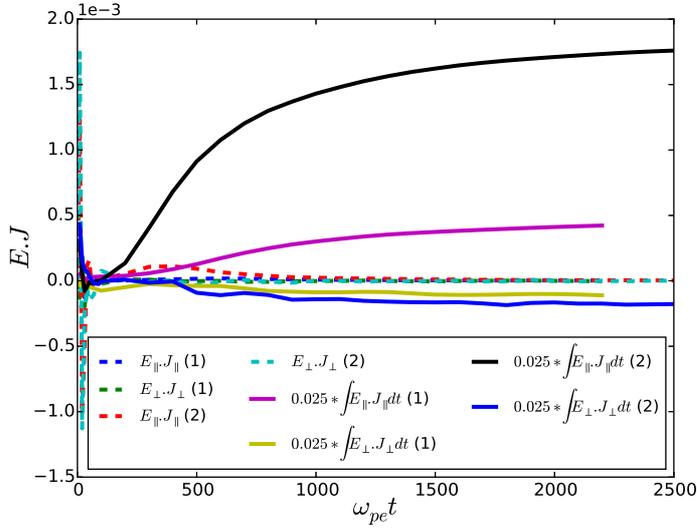}\hspace{0pc}%
\begin{minipage}[b]{13pc}\caption{\label{edotj}Time evolution of various $\bf{E}\cdot\bf{J}$ quantities. Parenthese (1) refers to the $\beta=0.092$ case, while parentheses (2) refers to $\beta=0.026$ case. The time integrals $\int{\bf{E}_{\parallel}\cdot\bf{J}_{\parallel}}dt$ and $\int{\bf{E}_{\perp}\cdot\bf{J}_{\perp}}dt$ are carried out over arithmetic average of $\bf{E}_{\parallel}\cdot\bf{J}_{\parallel}$ and $\bf{E}_{\perp}\cdot\bf{J}_{\perp}$ over the entire simulation box. The units of these integrals are arbitrary, multiplied by 0.025 to plot them in same scale. Strong fluctuations are seen in the $\bf{E}\cdot\bf{J}$ terms initially for $\omega_{pe} t<100$, which however do not contribute significantly to the time integral.}
\end{minipage}
\end{figure}
We next look at the term $\mathbf{E}\cdot\mathbf{J}$ in Fig.~\ref{edotj} to find the cause of energy dissipation and the corresponding particle heating. We split it up into components parallel to the local magnetic field ($\mathbf{E}_{\parallel}\cdot\mathbf{J}_{\parallel}$) and perpendicular to the field ($\mathbf{E}_{\perp}\cdot\mathbf{J}_{\perp}$). The time evolution of these quantities averaged over the simulation box are shown in Fig.~\ref{edotj}. A positive $\bf{E}\cdot\bf{J}$ value signifies energy dissipation. We find that initially, within the first $\omega_{pe}t=100$ time, the $\mathbf{E}\cdot\mathbf{J}$ term is quite noisy, developing large fluctuations. However after $\omega_{pe}t=100$ the term settles down and behaves smoothly. We believe the initial perturbations are due to the PIC simulation adjusting to the initial conditions which are not an exact eigen-solution to the PIC dispersion relation. We also plot the time integration of these terms, $\int {\bf{E}_{\parallel}\cdot\mathbf{J}_{\parallel}}dt$ and $\int {\bf{E}_{\perp}\cdot\mathbf{J}_{\perp}}dt$ in Fig.~\ref{edotj}. It does not show any strong growth during the first $\omega_{pe}t=100$ time. Also, if we look at the energy decay in these simulations in Fig.~\ref{energy}, no significant amount of energy is dissipated during this initial time. All this indicates that these initial fluctuations are not affecting energy dissipation in a significant way. We see that $\mathbf{E}_{\parallel}\cdot\mathbf{J}_{\parallel}$ is greater than $\mathbf{E}_{\perp}\cdot\mathbf{J}_{\perp}$ by a factor of 3-4 in both cases. The $\bf{E}\cdot\bf{J}$ terms are much stronger in the lower beta case. The time integrated plot shows this clearly, where $\int\mathbf{E}_{\parallel}\cdot\mathbf{J}_{\parallel}dt$ is stronger in the lower beta case. Furthermore, the $\bf{E}_{\parallel}\cdot\bf{J}_{\parallel}$ is positive, whereas $\bf{E}_{\perp}\cdot\bf{J}_{\perp}$ is negative. This analysis shows that $\mathbf{E}_{\parallel}\cdot\mathbf{J}_{\parallel}$ is the important term causing dissipation and heating of particles.

\begin{figure}[h]
\center
\includegraphics[width=0.9\textwidth]{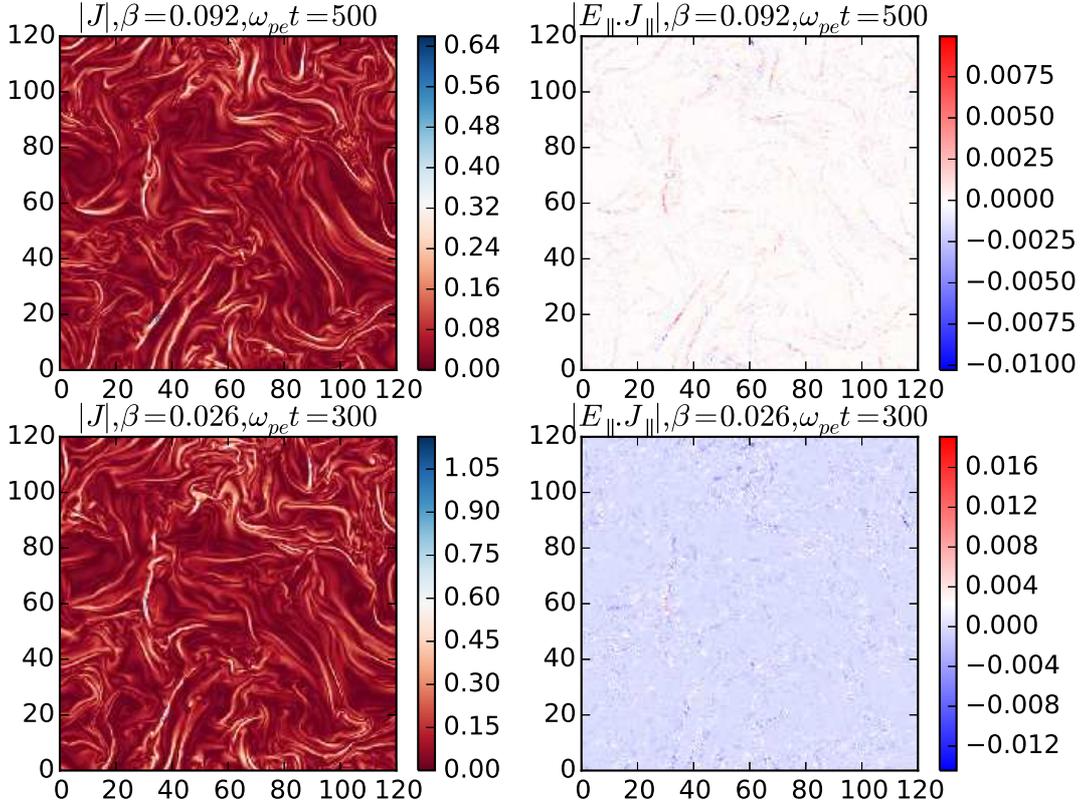}\hspace{2pc}%
\caption{\label{4plots}Cross-section color plots of $|\bf{J}|$ and $\bf{E}_{\parallel}\cdot\bf{J}_{\parallel}$ for the two cases at the specified times. The cross-section is taken perpendicular to the background magnetic field. The box dimensions are in units of skin-depth.}
\end{figure}
Next we look at the spatial profiles of these heating terms. In Fig.~\ref{4plots} we show the magnitude of current density $|\bf{J}|$ and $\bf{E}_{\parallel}\cdot\bf{J}_{\parallel}$ in a cross section of the simulation box perpendicular to the mean field. In case of $\beta=0.092$ the snapshot is taken at $\omega_{pe}t=500$, which corresponds to $0.36\tau_{A}$, where $\tau_A$ is the Alfven crossing time. From Fig.~\ref{energy} we see strong particle heating is taking place at this time for case 1. In case of $\beta=0.026$ the snapshot is taken at $\omega_{pe}t=300$, which also corresponds to $0.36\tau_A$. Again we see in Fig.~\ref{energy} strong particle heating at this time for case 2. As a result, we also see similar structure of current sheets in both the cases. The current sheets are very thin, of the order of skin depth $d_i$, in accordance with previous results~\cite{makwana2015}. The color scale shows that the intensity of current sheets is higher for the lower $\beta$ case. The $\bf{E}_{\parallel}\cdot\bf{J}_{\parallel}$ term is extremely localized in narrow regions in Fig.~\ref{4plots} for both the cases. These regions are well-correlated with the regions of strong current sheets. This indicates that the dissipation is extremely localized and occurring mostly in the current sheet regions. The dissipation term also looks stronger in the lower $\beta$ case.

\begin{figure}[h]
\center
\includegraphics[width=1.0\textwidth]{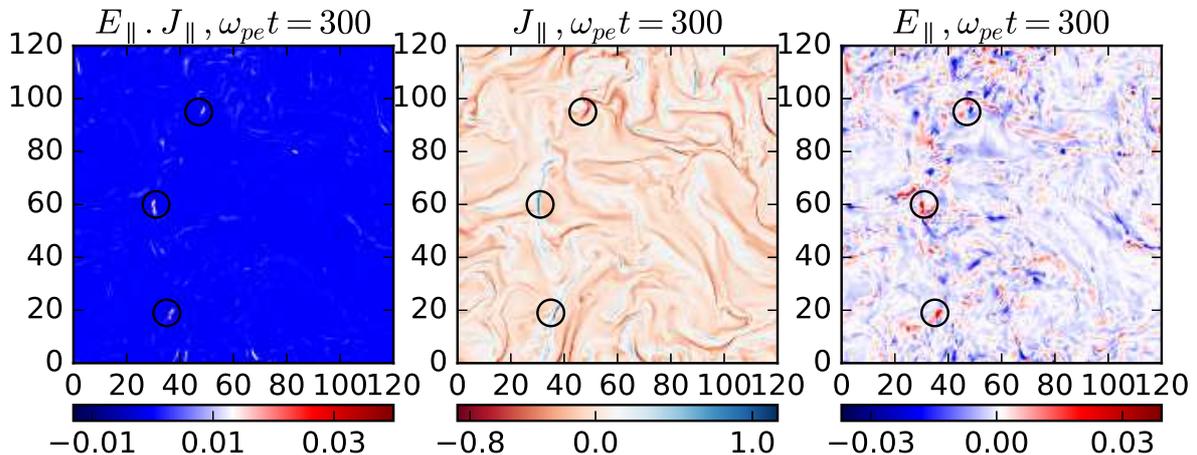}%
\caption{\label{e_j_ej_par} $E_{\parallel}.J_{\parallel}$, $J_{\parallel}$, and $E_{\parallel}$ (from left to right) in a plane perpendicular to the background magnetic field for the $\beta=0.026$ case at $\omega_{pe}t=300$. This plane is slightly shifted from the plane selected for Fig.~\ref{4plots}. The circles are highlighting three regions of high dissipation ($E_{\parallel}.J_{\parallel}$).}
\end{figure}

The parallel electric field $\bf{E}_{\parallel}$ is a necessary condition for magnetic reconnection~\cite{SchindlerHesse1988}, and often taken as an indicator of reconnection. Is the dissipation we observe in current sheets related to reconnection? We show the $\bf{E}_{\parallel}\cdot\bf{J}_{\parallel}$, $\bf{J}_{\parallel}$, and $\bf{E}_{\parallel}$ cross sections in a plane perpendicular to the background magnetic field in Fig.~\ref{e_j_ej_par} respectively for the lower $\beta$ case at $\omega_{pe}t=300$. The regions of high $\bf{E}_{\parallel}\cdot\bf{J}_{\parallel}$ are shown by putting circles around them. In Fig.~\ref{4plots} we saw that regions of high dissipation were correlated with regions of high total current density $|\bf{J}|$. Here we see that the current density is dominated by the parallel current density $J_{\parallel}$, and hence regions of strong dissipation are also correlated with strong parallel current. In the third plot of Fig.~\ref{e_j_ej_par}, we see that these circles of strong $\bf{E}_{\parallel}\cdot\bf{J}_{\parallel}$ are also regions of strong $\bf{E}_{\parallel}$. This indicates that reconnection in turbulent current sheets may be the cause of dissipation in this low-$\beta$ turbulence simulation. It should be noted that there are several regions with high $\bf{E}_{\parallel}$, but they are not corresponding to regions of high current density, thereby not giving strong dissipation.

\section{Discussion}
We have extended our previous work of kinetic simulations of decaying plasma turbulence at moderate plasma $\beta$ to low plasma $\beta$. We find that the decay rate of turbulence remains similar for this range of plasma $\beta$. Consequently, the heating rate of particles is also similar. The magnetic energy perpendicular wavenumber spectrum shows a slope of $k_{\perp}^{-1.3}$, which is similar to results at moderate plasma $\beta$. The kinetic energy spectrum does not show a clear inertial range for the lower $\beta$ simulations. At low plasma $\beta$ we also observe strong fluctuations in the electric fields at the beginning of the simulations. At low $\beta$ the Alfven velocity becomes relativistic, as do the currents and velocities. Also, the waveforms we use to setup waves in the initial conditions are derived from MHD, which are not eigen-functions of the kinetic system. We intend to remove these irregularities from our simulations by utilizing proper wave-dispersion relations in future work. 

Nevertheless, it appears that these strong electric fields die down quickly and do not affect the dissipation process significantly. In both moderate and low $\beta$ simulations, the parallel dissipation term, $\bf{E}_{\parallel}\cdot\bf{J}_{\parallel}$ is stronger than the perpendicular term, $\bf{E}_{\perp}\cdot\bf{J}_{\perp}$. Regions of strong dissipation are spatially well-correlated with regions of high current density sheets. In the low $\beta$ case they also correlate well with regions of strong parallel electric field, suggesting that reconnection in turbulent current sheets is playing a major role in this energy dissipation. Also in the low $\beta$ case, the particle heating results in the development of a non-thermal feature in the particle kinetic energy distribution function, with a slope close to $(\gamma-1)^{-1}$. However, recent simulations of reconnection also show non-thermal features which can be explained by two Maxwellians at different temperatures, the lower temperature Maxwellian for non-reconnected particles and a higher temperature Maxwellian for particles that have passed through the reconnection region. This aspect will need to be analyzed further in future work.

\ack{This work was supported in part by the National Science Foundation sponsored Center for Magnetic Self Organization (CMSO) at the University of Chicago, and by the Interuniversity Attraction Poles
Programme (initiated by the Belgian Science Policy Office, IAP P7/08 CHARM) and by the KU Leuven GOA/2015-014. This research used resources of the Bluewaters supercomputer at the National Center for Supercomputer Applications at the University of Illinois at Urbana-Champaign and of the National Energy Research Scientific Computing Center, a DOE Office of Science User Facility supported by the Office of Science of the U.S. Department of Energy under Contract No. DE-AC02-05CH11231.}

\end{document}